\documentclass[useAMS,usenatbib]{mn2e}

\usepackage{amssymb}
\usepackage{amsmath}
\usepackage[dvips]{graphicx}
\usepackage{epsfig}
\usepackage{natbib}
\usepackage{multicol}
\usepackage{rotating}
\usepackage{booktabs}
\usepackage{caption}
\usepackage{tikz}
\usetikzlibrary{shapes,arrows}

\label{firstpage}
\date{Accepted ... Received ...; in original form ...}
\pagerange{\pageref{firstpage}--\pageref{lastpage}} \pubyear{2012}
\title[Fits of a spherical model to GRB afterglows]{Applying an accurate spherical model to gamma-ray burst afterglow observations}
\author[Leventis et al.]{K. Leventis$^{1}$\thanks{E-mail: K.Leventis@uva.nl}, A. J. van der Horst$^{1}$, H. J. van Eerten$^{2}$, R.A.M.J. Wijers$^{1}$\\
$^{1}$Astronomical Institute `Anton Pannekoek', PO box 94248, 1090 SJ Amsterdam, the Netherlands\\
$^{2}$Center for Cosmology and Particle Physics, Physics Department, New York University, New York, NY 10003, USA}

\begin{document}

\maketitle

\begin{abstract}
We present results of model fits to afterglow data sets of GRB\,970508, GRB\,980703 and GRB\,070125, characterized by long and broadband coverage. The model assumes synchrotron radiation (including self-absorption) from a spherical adiabatic blast wave and consists of analytic flux prescriptions based on numerical results. For the first time it combines the accuracy of hydrodynamic simulations through different stages of the outflow dynamics with the flexibility of simple heuristic formulas. The prescriptions are especially geared towards accurate description of the dynamical transition of the outflow from relativistic to Newtonian velocities in an arbitrary power-law density environment. We show that the spherical model can accurately describe the data only in the case of GRB\,970508, for which we find a circumburst medium density $n\propto r^{-2}$. We investigate in detail the implied spectra and physical parameters of that burst. For the microphysics we show evidence for equipartition between the fraction of energy density carried by relativistic electrons and magnetic field. We also find that for the blast wave to be adiabatic, the fraction of electrons accelerated at the shock has to be smaller than $1$. We present best-fit parameters for the afterglows of all three bursts, including uncertainties in the parameters of GRB\,970508, and compare the inferred values to those obtained by different authors.

\end{abstract}


\vspace{1cm}


\section{Introduction}\label{intro}

Afterglow observations of gamma-ray bursts (GRBs) have provided important insight into the nature of these events. Some of it has been direct, for example the measurements of redshifts \citep{Metzger1997:Nat387}, or the association of some bursts with supernova explosions \citep{Hjorth2003:Natur423}. On the other hand, some has been indirect, accessible only once the available data are interpreted within the context of a physical model. The commonly used fireball model \citep{Rees1992,Paczynski1993}, for instance, is firmly supported by extensive modelling of afterglow observations as synchrotron radiation originating from a decelerating relativistic blast wave \citep{Wijers1997:mnras288,Waxman1997:ApJ489L, Sari1998:apjl497,Chevalier2000:ApJ536,Panaitescu2000:apj543}.

Despite the success of the aforementioned studies in interpreting afterglow observations within a general framework, the values derived by independent groups for the physical parameters of individual afterglows are often substantially different. Such is the case for the well-studied afterglows of GRB\,970508 and GRB\,980703, for which large differences can be found in the derived values for blast-wave energy, density of the circumburst medium (CBM) and microphysics parameters from different authors \citep{Wijers1997:mnras288,Granot2002:Apj568,Panaitescu2001:ApJ554, Panaitescu2002:ApJ571, Frail2003:apj590}. The CBM density seems to be especially unconstrained, as differences of many orders of magnitude can be found in the literature. One of the most important parameters of GRB outflows, that directly affects the inferred energetics and rate of these events, is the opening angle of the jet. Specifically, jetted instead of spherical outflows would significantly aleviate the energy requirements and boost the event rate of GRBs. The first strong inference of their presence \citep{Harrison1999:ApJ523L} was perceived as evidence for the ubiquitous role they play in the GRB phenomenon. Accumulating observations, however, have failed to fully confirm this picture, with many afterglows not showing any steepening in the light curves that can be attributed to a jet break (e.g. \citealt{Racusin2009:ApJ698}), rendering the influence of collimation on GRB outflows for the most part ambiguous. All these uncertainties on the inferred physical parameters of GRB blast-waves have called for refinement and greater precision in the methods that underlie afterglow modelling.

Theoretical afterglow calculations have been continuously improved to include more precise methods of calculating the dynamics and spectra of the source (e.g. \citealt{Kobayashi1999:apj513, Huang1999:mnras309, Rhoads1999:Apj525, Granot2002:Apj568, Granot2012:MNRAS421, Peer2012:ApJ752L}). Many recent studies (e.g. \citealt{Meliani2007:mnras376, Zhang2009:Apj698, vanEerten2010:mnras403, Wygoda2011:Apj738, DeColle2012:Apj746, vanEerten2012:Apj749}) are based on high-resolution relativistic hydrodynamic (RHD) simulations which are essential to understand critical aspects of the outflow's dynamics, like lateral spreading of jets and the transition to the non-relativistic phase. This allows in principle for accurate determination of spectra and light curves from simulation runs. However, this method is not suitable for iterative fitting of model parameters to observations due to the limitations posed by the necessary performance of numerous time-consuming RHD simulations.

Recently (\citealt{vanEerten2012:Apj749, Leventis2012:MNRAS427}; see also \citealt{vanEerten2012:apj747}) a new approach has been developed for the calculation of spectra and light curves that retains the accuracy of the numerical techniques, without requiring the long run times of simulations. While the methods of these studies differ, they are common in how they are based on sets of blast-wave simulations that span the parameter space. In the case of \cite{vanEerten2012:Apj749} dynamical results of 2D simulations have been tabulated allowing the user to perform a straightforward numerical calculation of the afterglow radiation for any combination of the physical parameters within the explored range. Even so, this calculation can be lengthy and is best executed on a parallel computer network.

The method of \cite{Leventis2012:MNRAS427} is based on 1D RHD simulations that span the entire range of dynamics, from ultrarelativistic to Newtonian velocities. These simulations, however, do not account for jet features as they rely on the assumption of spherical symmetry. Several runs have been used to calibrate analytically derived scalings of observed synchrotron spectra. The resulting formulas have the unique advantage of combining the accuracy of high-resolution trans-relativistic simulations with the versatility of analytic equations. The fact that they cover a sequence of dynamical phases has motivated us to use them in order to fit model parameters to observational data for afterglows with extensive monitoring. The bursts we are mainly concerned with in this paper are GRB\,970508, GRB\,980703 and GRB\,070125, all monitored in several bands from radio to X-ray frequencies and covering observer times from hours to several months. The two former are among the most studied afterglows with several groups publishing results they have obtained through afterglow modelling.

In this work we present fit results for the afterglows of these bursts and investigate the extent to which a spherical outflow can provide an adequate description of the data. These results also serve as a basis for comparison to model fits based on 2D simulations. Furthermore, the prescriptions of \cite{Leventis2012:MNRAS427} enable us to examine the density structure of the burster's immediate environment, as a continuous range of values for the slope of the CBM density is allowed. The resulting slope can then reveal unusual density distributions of the CBM, or confirm previous claims based on models with only preset structures available, typically constant density or a profile corresponding to a stellar-wind environment ($\propto \, r^{-2}$).

The paper is organized as follows. A description of the observational data that have been used during fitting is presented in Section \ref{data}. In Section \ref{model} we illustrate the main features of the physical model we have used and in Section \ref{results} we present our main results. In Section \ref{970508_section} we focus on the inferred parameters of GRB\,970508 for which we obtain the most reliable results. In Section \ref{discussion} we discuss the implications of this work on afterglow physics and modelling. Finally, in Section \ref{conclusions} we conclude by summarizing our main findings.


\section{Data}\label{data}

In this study we focus on three sources: GRB\,970508, GRB\,980703 and GRB\,070125. 
All three have well-sampled afterglows across the electromagnetic spectrum. In particular they are among the few GRBs that have detections in multiple radio bands at hundreds of days after the initial gamma-ray trigger. This allows us to model the full evolution of the GRB blast wave from the ultrarelativistic to the non-relativistic phase. Another burst with afterglow monitoring spanning almost a decade in the radio is GRB\,030329. We have not fit that data set as it is clear from the light curves that a jetted model is needed to interpret the observations (see \citealt{vanderHorst2008:aap480} and references therein).

Since the lauch of the \textit{Swift} satellite, it has become clear that the early ($10^3-10^5\, \textrm{s}$) afterglow behaviour of many bursts cannot be explained by standard afterglow models \citep{Nousek2006:ApJ642}. Energy injection into the blast wave has been proposed to explain the typically shallow decay that the optical and X-ray light curves show (e.g. \citealt{Granot2006:MNRAS366L, Nousek2006:ApJ642, Zhang2006:ApJ642, Panaitescu2011:MNRAS414}). Other plausible explanations are evolution of the shock microphysics parameters \citep{Granot2006:MNRAS370}, or viewing angle effects \citep{Eichler2006:ApJ641L}. In our sample, GRB\,970508 and GRB\,070125 display an atypical behaviour, lasting in both cases up to $1.5$ days. Especially for GRB\,970508, the fast-rising optical light curves before $1.5$ days may reveal a refreshed shock, occuring when a slow shell catches up with the afterglow shock at later times \citep{Kumar2000:ApJ532, Granot2003:Natur426}. After $1.5$ days the light curves are compatible with the canonical afterglow decay. Processes like energy injection, refreshed shocks and effects due to off-axis viewing angle cannot be accounted for in the model we are using in this work. For this reason we have excluded data before $1.5$ days from the fitted data sets of both GRB\,970508 and GRB\,070125.

For GRB\,970508 radio observations were performed at $1.43$, $4.86$ and $8.46$ GHz \citep{galama1998:apj500,Frail2000:apj537}. 
Near-infrared and optical data have been published at 6 observing bands 
\citep{chary1998:apj498,galama1998:apj497,sokolov1998:aa334,sahu1997:apj489,garcia1998:apj500}. 
The magnitudes of the underlying host galaxy in the $B$, $V$, $R_\textrm{c}$ and $I_\textrm{c}$ bands have been presented in 
\citet{zharikov1999:aas138}, while the observations in the $K$ and $U$ bands are sufficiently early that they are not affected by the host galaxy brightness. 
We have corrected the observed optical magnitudes for galactic extinction, subtracted the host galaxy flux, and converted them to fluxes. 
The afterglow was observed in X rays with BeppoSAX \citep{piro1998:aa331}, 
for which we have converted the X-ray count rates to fluxes by assuming a spectral index of $-1.1$ over the observing band. 

GRB\,980703 was observed at the same radio frequencies as GRB\,970508 \citep{berger2001:apj560,Frail2003:apj590}. 
We have used all the near-infrared and optical data in the $H$, $J$, $I$, $R$, $V$ and $B$ bands 
\citep{bloom1998:apj508,castrotirado1999:apj511,vreeswijk1999:apj523}. 
We have corrected the observed magnitudes for galactic extinction, but also for extinction in the host galaxy with $E(B-V)=0.29$ \citep{Starling2007:apj661, Starling2008:A&A488}. 
The host galaxy of GRB\,980703 was bright, not only in the optical \citep{Frail2003:apj590} but also at radio wavelengths \citep{berger2001:apj560}, 
and we have subtracted the host galaxy flux at all these wavelengths from our measured fluxes. 
The afterglow has also been detected at X-ray energies \citep{vreeswijk1999:apj523}, 
for which we have used the same conversion method as in the case of GRB\,970508.

For GRB\,070125 we have used all the broadband data presented in \cite{DeCia2011:MNRAS418}. Radio observations were performed at $4.86$, $8.46$, $15$ and $22.5\, \textrm{GHz}$, while millimetre observations were carried out at $95$ and $250\, \textrm{GHz}$. The data set is supplemented by observations at 12 more bands ranging from the near infrared to X-ray energies, including optical and ultraviolet bands.

To carry out the modelling we have adopted the following cosmology: $\Omega_{\textrm{M}}=0.27$, $\Omega_{\Lambda}=0.73$ 
and the Hubble-parameter $H_0=71\,\mbox{km}\,\mbox{s}^{-1}\,\mbox{Mpc}^{-1}$; 
so for the GRB\,970508 redshift of $z=0.835$ \citep{Metzger1997:Nat387} the luminosity distance is $d_{\textrm{L}}=1.64\cdot 10^{28}\mbox{cm}$, for GRB\,980703 the redshift $z=0.966$ \citep{Djorgovski1998:ApJ508L} corresponds to $d_{\textrm{L}}=1.96\cdot 10^{28}\mbox{cm}$, and for GRB\,070125 the redshift $z=1.547$ \citep{Cenko2008:ApJ677} implies that $d_{\textrm{L}}=3.53\cdot 10^{28}\mbox{cm}$. 

During the fit process, no data were excluded based on flux values. However, in the figures we present, data that are not significant at the $2\sigma$ level are depicted as upper limits, for display purposes.

\subsection{Scatter of the data}\label{scatter}
A noticeable feature of radio data is the high degree of scatter they show, especially compared to the size of the error bars (see Section \ref{results}). In the case of GRB\,970508, notable scatter is also present in near-infrared and optical frequencies. In these bands it is presumably caused by the use of data from various telescopes without the performance of cross-calibration analysis.

In the radio, interstellar scintillation affects the flux levels \citep{Goodman1997:NewA2}. Its strength diminishes as the angular size of the source grows and this has been used to infer the radius of GRB outflows \citep{Frail1997:Nat389, Taylor1997:Natur389, Waxman1998:ApJ497, Frail2000:apj537, Yost2003:apj597}. Various other groups have accounted for the effect of scintillation \citep{Panaitescu2002:ApJ571, Chandra2008:ApJ683}, especially in early data, by effectively increasing the size of the error bars, more so in early observer times. In our study we have not included the effect of scintillation in the model. One reason is that it does not affect the best-fit values of our results significantly, as the central values of the measurements are not perturbed. Another reason is the lack of detailed measurements for the amount of scattering material off the galactic plane, which makes the effect of scintillation in models of extragalactic sources uncertain \citep{Chandra2008:ApJ683}.


\section{The model}\label{model}

\subsection{General description}

The model we have used is a direct implementation of the method presented in \cite{Leventis2012:MNRAS427}. In that paper we present simulation-calibrated flux prescriptions of synchrotron radiation, including self-absorption, throughout the entire dynamical evolution of GRB afterglows. The model assumes an initially ultrarelativistic spherical blast wave expanding adiabatically inside a medium with a density profile described by a power law: $n(r) \! \propto \! r^{-k}$. The energy distribution of the electrons accelerated at the forward shock is also assumed to be a power law. The minimum Lorentz factor of that distribution is calculated through the energy density and mass density of the shocked gas. The synchrotron spectrum is then determined through the emissivity and absorption coefficient of these relativistic electrons.

In total there are seven free parameters. These are the blast-wave energy $E_{52}$ in units of $10^{52}\,\textrm{erg}$, the number density $n_0$ at $10^{17}\,\textrm{cm}$ (regardless of the density structure), the index $p$ of the electron power-law distribution, the index $k$ of the density distribution of the matter surrounding the GRB, the fraction $\xi$ of accelerated electrons, and $\epsilon_{\textrm{e}}$ and $\epsilon_{\textrm{B}}$ denoting the fractions of internal energy carried by the relativistic electrons and magnetic field, respectively. In practice, due to a degeneracy of this model (\citealt{Eichler2005:apj627}) a value for one of these parameters has to be assumed in order to uniquely determine the others. In this work we `break' the degeneracy by assuming $\xi=1$ in all runs, unless otherwise stated.

The flux prescriptions are based on analytic calculations of flux scalings during the relativistic \citep{Blandford1976} and Newtonian \citep{Sedov1959, Taylor1950RSPSA201} phase of the blast-wave dynamics. In these two dynamical regimes the flux at every power-law segment of the spectrum has been calibrated in terms of $p$ and $k$. Several hydrodynamic simulations of the afterglow dynamics were run and subsequently post-processed using a radiative-transfer code \citep{vanEerten2009:mnras394, vanEerten2010:mnras403}. The calibration was carried out by matching analytic expressions for the flux scalings to these numerical results. The sharpness of spectral breaks connecting different power laws of the spectrum is also expressed as a function of $p$ and $k$. The transition from the relativistic to the Newtonian solution is nicely described as a temporal power-law break between the asymptotic behaviour of the critical parameters of the spectrum, namely maximum flux $F_{\textrm{m}}$, self-absorption frequency $\nu_{\textrm{a}}$ and synchrotron characteristic frequency of the lowest-energy electrons $\nu_{\textrm{m}}$. It is worth noting that the characteristics (break time and sharpness) of those temporal breaks are, in general, unique for every parameter of the spectrum. This emphasizes the advantages of simulation-based flux prescriptions compared to simple analytic models for the transrelativistic behaviour of observed afterglows.

\subsection{The cooling break}

A feature of the synchrotron spectrum not covered in the treatment of \citet{Leventis2012:MNRAS427} is the cooling break, manifested as a fourth spectral parameter $\nu_{\textrm{c}}$. Its presence in the spectrum, however, might be important, especially for observations at optical wavelengths and X-ray energies. For that reason all the performed fits have been checked for consistency by calculating the value of $\nu_{\textrm{c}}$ according to formulas available in the literature (e.g. \citealt{Granot2002:Apj568,vanEerten2009:mnras394}) and comparing it to the frequencies of the observations. The results of the two aforementioned studies are compatible. We have chosen to use those of \citet{vanEerten2009:mnras394} due to the fact that a general value for $k$ is allowed in their prescriptions. The consistency checks have been performed throughout the range of observer times covered by the data. A value of $\nu_{\textrm{c}}$ greater than the observing frequencies implies that cooling has not affected the fits and the obtained values for the physical parameters are consistent with the underlying physical model. To the best of our knowledge simulation-based analytic prescriptions for $\nu_{\textrm{c}}$ beyond the relativistic phase do not exist in the literature. That being the case, we have used formulas applicable in this phase throughout. This extrapolation provides a lower limit on the actual value of $\nu_{\textrm{c}}$ because its temporal slope in the Newtonian phase is shallower than in the relativistic (\citealt{vanEerten2010:mnras403}), which is sufficient when $\nu_{\textrm{c}}$ is found not to interfere with the observing frequencies.

On the other hand, when the value of $\nu_{\textrm{c}}$ is found to be lower than -- or at about the same levels as -- the observing frequencies a different approach is necessary in order to firmly constrain the influence of cooling on the data. Our fitting code has been expanded to include a prescription for the position of $\nu_{\textrm{c}}$ as a function of time. We have made use of the formulas from \citet{vanEerten2009:mnras394} by calculating $\nu_{\textrm{c},1}$ of that paper. Formally this expression should only apply in the case of slow cooling ($\nu_{\textrm{m}} < \nu_{\textrm{c}}$). However, it is easy to verify (see also \citealt{Granot2002:Apj568}) that the expression for $\nu_{\textrm{c}}$ in the case of fast cooling gives a similar result within a factor of about 2. A few modifications in the prescriptions are then required in order to account for the influence of cooling in the broadband spectrum. When $\nu_{\textrm{a}} < \nu_{\textrm{m}} < \nu_{\textrm{c}}$ or $\nu_{\textrm{m}} < \nu_{\textrm{a}} < \nu_{\textrm{c}}$ the only modification is that of appending another break in the spectrum at the cooling frequency, beyond which the spectrum steepens by a half (\citealt{Sari1998:apjl497}). The formula we have used is
\begin{eqnarray}\label{eq-spectrum}
F_{\nu}(\nu_{\textrm{obs}}) =  A\, \left[ \left( \frac{\nu_{\textrm{obs}}}{\nu_0} \right)^{-a_1 \, s} + \left( \frac{\nu_{\textrm{obs}}}{\nu_0} \right)^{-a_2 \, s} \right] ^{-1/s} \,\, \times \nonumber \\
 \left[ 1+\! \left( \frac{\nu_{\textrm{obs}}}{\nu_1} \right)^{\! h(a_2-a_3)} \right]^{-1/h}\!\! \times \left[ 1+\! \left( \frac{\nu_{\textrm{obs}}}{\nu_2} \right)^{\! r(a_3-a_4)} \right]^{-1/r}
\end{eqnarray}
The first line in eq. (\ref{eq-spectrum}) describes the first break of the spectrum at the lowest characteristic frequency, while each factor on the second line stands for an extra break at progressively higher frequencies. The parameters $\nu_0,\, \nu_1,\, \nu_2$ and $s,\, h,\, r$ represent the values of the three critical frequencies and the sharpness of the spectral breaks they correspond to, respectively, while $a_1,\, a_2,\, a_3$ and $a_4$ are the slopes of the four power laws present in a spectrum with three breaks. Finally, $A$ is the normalising factor of the spectrum derived through modelling of the peak flux $F_{\textrm{m}}$.

When $\nu_{\textrm{m}},\, \nu_{\textrm{c}} < \nu_{\textrm{a}}$ the ordering of $\nu_{\textrm{m}}$ and $\nu_{\textrm{c}}$ does not play a role and one retrieves spectrum 3 of \citet{Granot2002:Apj568}. In that case we have approximated the self-absorption frequency with the values applicable to the no-cooling case. Similarly, when $\nu_{\textrm{a}} < \nu_{\textrm{c}} < \nu_{\textrm{m}}$ (spectrum 5 of \citealt{Granot2002:Apj568}) we have approximated both $\nu_{\textrm{m}}$ and $\nu_{\textrm{a}}$ with their values in the absence of cooling, while the peak flux is attributed to $\nu_{\textrm{c}}$. Formally, when $\nu_{\textrm{a}} < \nu_{\textrm{c}} < \nu_{\textrm{m}}$ the self-absorption break is split in two break frequencies with an extra power-law segment between them that has a slope of $11/8$. We have neglected that effect and used only one self-absorption frequency that has the value of $\nu_{\textrm{a1}}$ from \cite{Leventis2012:MNRAS427}. This frequency connects power laws of slope $2$ and $1/3$. In reality, we have found that most of the time best-fit values of the physical parameters imply that these approximations are not used since $\nu_{\textrm{c}} > \nu_{\textrm{a}},\, \nu_{\textrm{m}}$. However there are instances when this is not the case and we address these in more detail in Section \ref{970508_section}.

A last issue that needs to be dealt with when cooling influences the fits is the application of the relativistic formulas for $\nu_{\textrm{c}}$ throughout the range of observer times. To assess the validity of this application one needs to estimate the duration of the relativistic phase of the afterglow in the observer frame. In the absence of a detailed description for the transrelativistic behaviour of the cooling frequency, the most general way to do that is by calculating the observer time which corresponds to the transition between the relativistic and Newtonian asymptotes, $t_{\textrm{NR}}$ (e.g. \citealt{Piran2004, Leventis2012:MNRAS427}). This calculation has been performed for all sets of best-fit parameters and is presented along with our main results in Section \ref{results}.

\subsection{Fitting procedure}

The fitting method we have used is a $\chi^2$-minimization algorithm following the downhill-simplex method combined with simulated annealing, as explained in \cite{vanEerten2012:Apj749}. The errors for the best-fit parameters of GRB\,970508 have been determined via a Monte Carlo process. In this analysis the values of all data points are perturbed randomly, based on their error bars, and a new best-fit set of parameters is calculated for the synthetic data. For every physical scenario (class and constraint) this has been repeated $1000$ times from which $683$ best fits were drawn to determine the range of the parameters' values at a $68.3\%$, i.e. $1\sigma$, confidence level.

The fitted parameters were allowed to vary within the following ranges ($n_0$ in \textit{cgs} units): $10^{-5}<E_{52}<10^{4}$, $10^{-5}<n_0<10^5$, $2.0<p<3.5$, $10^{-7}<\epsilon_{\textrm{B}}<1.0$, $10^{-5}<\epsilon_{\textrm{e}}<1.0$, $-0.5<k<2.5$.


\begin{figure*}
\centering
\includegraphics[trim=0cm 8cm 0cm 0cm, clip=true, width=2.0\columnwidth]{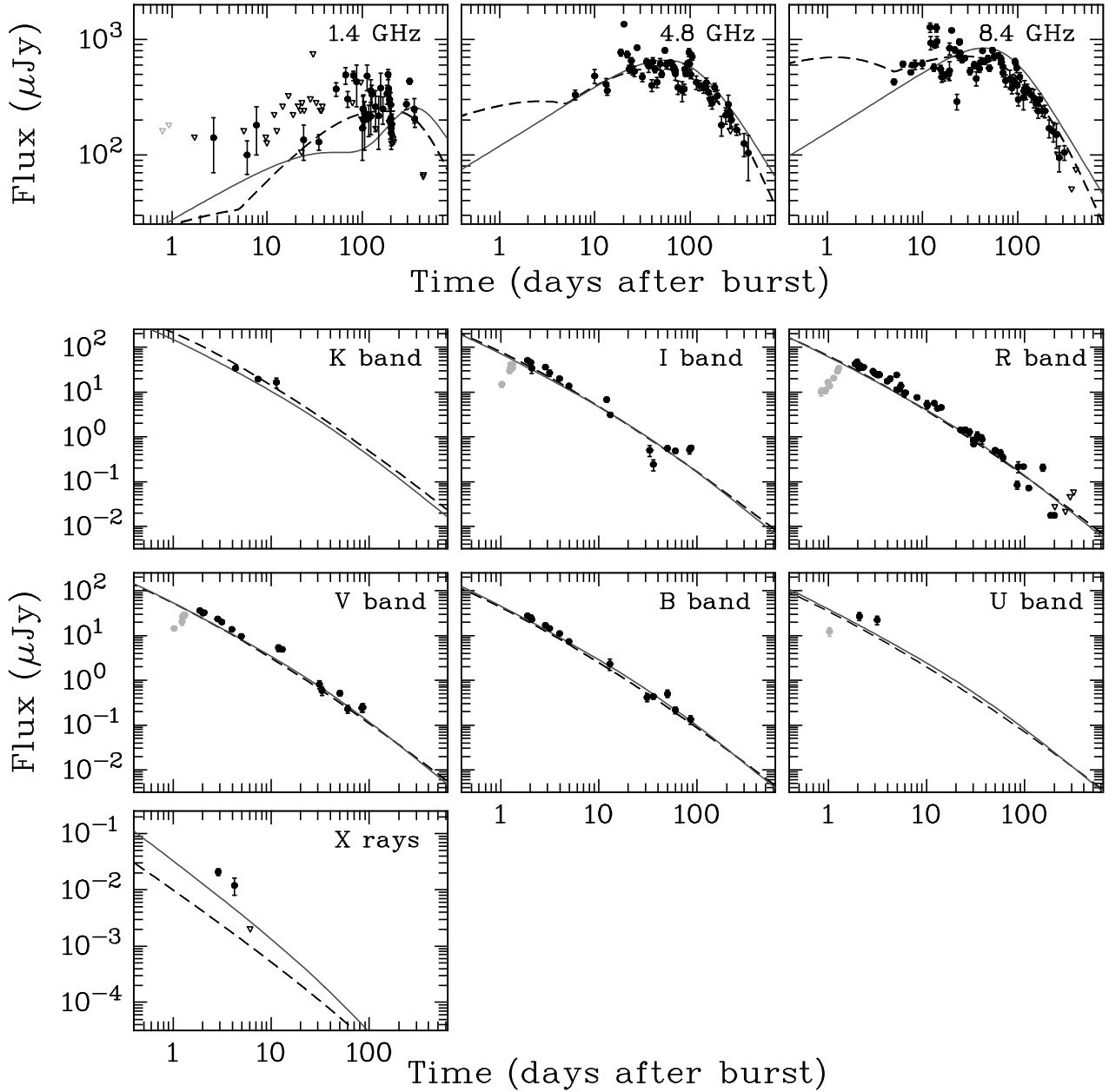}
\parbox{2.0\columnwidth}{\caption{Afterglow of GRB\,970508. Best-fit light curves for \textit{ISM} (solid grey line) and \textit{Wind} (dashed black line) classes. When $k$ is a free parameter, the \textit{Wind} scenario is retrieved with high precision. The three radio bands are on top and the rest follow in order of increasing frequency, spanning near-infrared, optical, ultraviolet and X-ray energies. Data points before $1.5\, \textrm{days}$ have been excluded from the fits but appear in grey in the figure. In all bands, data points have $1\sigma$ errors. Triangles depict upper limits at the $2\sigma$ level. \label{970508_lcs}}}
\end{figure*}

\begin{table*}
\centering  
\parbox{2.05\columnwidth}{\captionof{table}{Best-fit parameters, with $1\sigma$ errors, for GRB\,970508 in all classes of models and for all microphysics settings. The fits have been performed to data including radio, near-infrared, optical, ultraviolet and X rays, but excluding observations made prior to $1.5$ days after the gamma-ray trigger (see Section \ref{data}). The uncertainties in the last row have been calculated for 5 times higher error bars. The third column contains the blast-wave energy in units of $10^{52}\,\textrm{erg}$. The fourth column represents $n_0$, the number density at a radius of $10^{17} \, \textrm{cm}$. When $k=2$, $n_0$ and $A_{\ast}$ \citep{Chevalier2000:ApJ536} are related by the formula $n_0\simeq 30\, A_{\ast}$. For example, the best-fit model (equipartition constraint) of the \textit{Wind} class has $A_{\ast}=0.243\, \textrm{g}\,\textrm{cm}^{-1}$. The last column presents the value of $\chi^2$ divided by the degrees of freedom (dof). \label{970508_table}}}
\small\addtolength{\tabcolsep}{+0.7pt}
\begin{tabular}{l c c c c c c c c} 
\toprule
 $\textrm{Class}$ & $\textrm{Constraint}$& $E_{52}$ & $n_0$ & $p$ & $\epsilon_{\textrm{B}}$ & $\epsilon_{\textrm{e}}$ & $k$ & $\chi^2/\textrm{dof}$ \\ [0.5ex] 
\hline
\hline
$$ & $-$ & $1.227^{+0.911}_{-0.360}$ & $15.02^{+25.25}_{-6.62}$ & $2.483^{+0.024}_{-0.022}$ & $(4.2^{+11.3}_{-3.7})\!\cdot\!10^{-4}$ & $0.702^{+0.298}_{-0.122}$ & $0$ & $\textbf{29.89}$ \\ [1.5ex]
$\textit{ISM}$ & $\textrm{Equipartition}$ & $0.300^{+0.047}_{-0.033}$ & $0.319^{+0.098}_{-0.084}$ & $2.279^{+0.020}_{-0.019}$ & $0.337^{+0.023}_{-0.026}$ & $0.337^{+0.023}_{-0.026}$ & $0$ & $40.05$ \\ [1.5ex]
$$ & $\textrm{Medvedev}$ & $0.333^{+0.024}_{-0.020}$ & $0.784^{+0.111}_{-0.133}$ & $2.307^{+0.013}_{-0.011}$ & $0.130^{+0.009}_{-0.010}$ & $0.361^{+0.013}_{-0.014}$ & $0$ & $35.63$  \\ [0.5ex]
\hline
$$ & $-$ & $0.131^{+0.006}_{-0.045}$ & $7.516^{+2.934}_{-2.080}$ & $2.277^{+0.017}_{-0.053}$ & $0.551^{+0.449}_{-0.092}$ & $0.589^{+0.411}_{-0.045}$ & $2$ & $28.55$ \\ [1.5ex]
$\textit{Wind}$ & $\textrm{Equipartition}$ & $0.134^{+0.003}_{-0.007}$ & $7.263^{+0.133}_{-0.258}$ & $2.280^{+0.014}_{-0.013}$ & $0.575^{+0.028}_{-0.014}$ & $0.575^{+0.028}_{-0.014}$ & $2$ & $\textbf{28.46}$ \\ [1.5ex]
$$ & $\textrm{Medvedev}$ & $0.121^{+0.009}_{-0.005}$ & $8.858^{+0.150}_{-0.408}$ & $2.259^{+0.015}_{-0.011}$ & $0.448^{+0.038}_{-0.054}$ & $0.669^{+0.028}_{-0.042}$ & $2$ & $28.56$  \\ [0.5ex]
\hline
$$ & $-$ & $0.131^{+0.857}_{-0.006}$ & $7.465^{+19.890}_{-0.491}$ & $2.277^{+0.240}_{-0.017}$ & $0.555^{+0.042}_{-0.555}$ & $0.595^{+0.405}_{-0.038}$ & $1.983^{+0.046}_{-2.483}$ & $28.64$ \\ [1.5ex]
$k\,\textit{free}$ & $\textrm{Equipartition}$ & $0.133^{+0.003}_{-0.004}$ & $7.207^{+0.267}_{-0.223}$ & $2.280^{+0.015}_{-0.013}$ & $0.580^{+0.016}_{-0.028}$ & $0.580^{+0.016}_{-0.028}$ & $1.983^{+0.047}_{-0.016}$ & $\textbf{28.54}$ \\ [1.5ex]
$$ & $\textrm{Medvedev}$ & $0.122^{+0.014}_{-0.003}$ & $8.717^{+0.323}_{-0.374}$ & $2.260^{+0.014}_{-0.013}$ & $0.453^{+0.020}_{-0.073}$ & $0.673^{+0.015}_{-0.057}$ & $1.972^{+0.046}_{-0.036}$ & $28.64$  \\ [0.5ex]
\bottomrule\\ [-1.0ex]
$k\,\textit{free}$ & $\textrm{Equipartition}$ & $0.133^{+0.130}_{-0.065}$ & $7.207^{+14.088}_{-3.156}$ & $2.280^{+0.067}_{-0.243}$ & $0.580^{+0.420}_{-0.237}$ & $0.580^{+0.420}_{-0.237}$ & $1.983^{+0.517}_{-0.389}$ & $1.158^{\,a}$ \\ [1.0ex]
\bottomrule \\ [-1.0ex]
\multicolumn{6}{p{0.5\textwidth}}{$^{a\,}$Error bars of data points are rescaled by a factor of $5$.}
\end{tabular}
\smallskip
\end{table*}

\section{Results}\label{results}

For all afterglow data sets, we present three classes of models. Each class corresponds to a different assumption (or the lack thereof) for the value of $k$. We have run fits for $k=0\textrm{ and }2$, corresponding to constant density CBM (labelled \textit{ISM}) and a constant-stellar-wind profile (labelled \textit{Wind}), respectively, and fits where $k$ is a free parameter. For each class, a range of microphysics settings has been tested. Namely, we have either allowed for both $\epsilon_{\textrm{e}}$ and $\epsilon_{\textrm{B}}$ to be free parameters, or connected them through a closure relation that effectively reduces them to one free parameter. Two options for the closure relation have been explored. On the one hand we have imposed equipartition ($\epsilon_{\textrm{e}}=\epsilon_{\textrm{B}}$) and on the other the `Medvedev' relation ($\epsilon_{\textrm{e}}^2 = \epsilon_{\textrm{B}}$; \citealt{Medvedev2006:apj651}). All other parameters have been kept free at all runs, apart from $\xi$ which, for every run, has taken the value of 1.

\subsection{GRB\,970508}\label{970508_sub}
We have performed several fits both to the full data set and to different subsets (radio only, radio and optical only, radio, optical and X rays) of the afterglow observations of GRB\,970508. Radio data alone do not provide enough information to determine simultaneously all the parameters. However, when $k$ is frozen (either in the \textit{ISM} or the \textit{Wind} scenario) and a microphysics constraint is used, the results from fitting the radio only, are fairly similar to those from fits to the full data set; all best-fit values of parameters are less than $50\%$ off in the \textit{Wind} class and less than a factor of $2$ off in the \textit{ISM} class. Including X-ray data has almost no influence on the inferred values of the physical parameters, as the fits are governed by the combination of radio and optical observations. Nevertheless, we present results and light curves from fits to all bands for completeness.

In Fig. \ref{970508_lcs} we present light curves of best-fit models applied to the full data set. We have found that the spherical model can produce an adequate fit to the data, when $k=2$. Results for the \textit{Wind} scenario are almost identical to those from fits where $k$ is a free parameter. Models of the \textit{ISM} class consistently overpredict late radio flux at $4.86$ and $8.46\,\textrm{GHz}$. On the other hand, \textit{Wind} models provide a good description at all observer times. In the optical and near-infrared bands, the \textit{ISM} and \textit{Wind} cases are practically indistinguishable. One common feature of both is the systematic, albeit minor, underprediction of early ($<10\,\textrm{days}$) flux, especially in the $R$ and $V$ bands. This is less pronounced in the surrounding $K$, $I$ and $B$ bands. It is worth noting that the X-ray data cannot be fitted by any combination of parameters. Along with the fact that the flux drops sharply after the first two data points, this hints towards a separate origin of the early X-ray flux, for example, inverse Compton (e.g. \citealt{Sari2001:ApJ548}). Alternatively, the high X-ray flux at early times could be due to flaring activity, which is not temporally resolved due to the poor coverage.

In the \textit{Wind} scenario, all critical frequencies lie below the near-infrared. On the other hand, both $\nu_{\textrm{a}}$ and $\nu_{\textrm{m}}$ pass through the radio bands. This is in rough agreement with the findings of \cite{Chevalier2000:ApJ536} and \cite{Panaitescu2002:ApJ571}, although we do not confirm the expectations of the former group regarding the passage of $\nu_{\textrm{c}}$ from the optical. Instead we find that $\nu_{\textrm{c}}$ stays below $10^{14}\,\textrm{Hz}$ during the observations. In the \textit{ISM} case we find that $\nu_{\textrm{m}}$ starts off between the optical and radio and crosses $\nu_{\textrm{a}}$ ($5\cdot10^{9}\,\textrm{Hz}$) at $\sim50$ days. We also find that $\nu_{\textrm{c}}$ remains between optical and X-ray energies throughout, contrary to the results of \cite{Galama1998:ApJ500L} and \cite{Wijers1999:apj523} who find that $\nu_{\textrm{c}}$ crosses the optical frequencies early on. Calculation of $t_{\textrm{NR}}$ yields $145$ and $180$ days, in the best-fit models of the \textit{ISM} and \textit{Wind} class, respectively.

In Table \ref{970508_table} we present best-fit parameters, with $1\sigma$ errors, of runs to the full data set. A readily apparent feature is the value of $k$ when it is a free parameter, which converges to the \textit{Wind} scenario. Actually, all best-fit values as well as the $\chi^2$ of these two classes are almost identical, regardless of the chosen microphysics. From the \textit{ISM} class only the run with no constraints on the microphysics comes close in terms of $\chi^2$, but that model requires a low value for $\epsilon_{\textrm{B}}$ and high value for $\epsilon_{\textrm{e}}$ to work. The energy inferred in this case is an order of magnitude higher than the values corresponding to the \textit{Wind} scenario.

\begin{figure*}
\centering
\includegraphics[trim=0cm 8cm 0cm 0cm, clip=true, width=2.0\columnwidth]{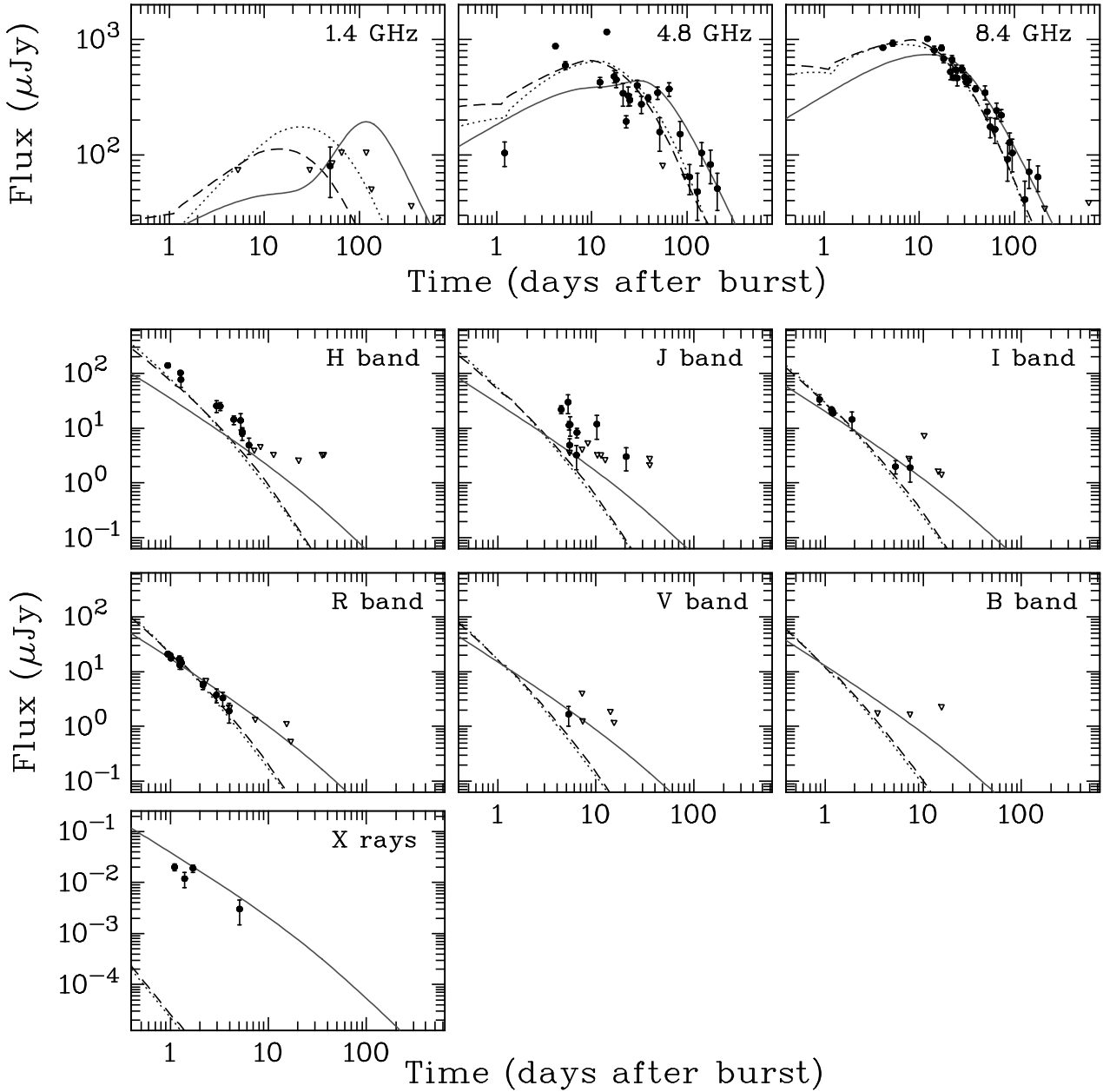}
\parbox{2.0\columnwidth}{\caption{Afterglow of GRB\,980703. Best-fit light curves for \textit{ISM} (solid grey line), \textit{Wind} (dotted black line) and \textit{k free} (dashed black line) classes. Radio bands are shown in the top panel. The lower panel contains near-infrared, optical and X-ray bands. All data were taken into account for the light curves we present. In all bands, data points have $1\sigma$ errors. Triangles depict upper limits at the $2\sigma$ level. \label{980703_lcs}}}
\end{figure*}

For all classes of models, the best-fit values of $\chi^2/\textrm{dof}$ are much higher than $1$. This is mainly caused by the notable scatter that data in radio, near-infrared and optical bands show. The scatter (discussed in Section \ref{scatter}) is not reflected in the size of the error bars. This is clearly demonstrated in the very small uncertainties that the inferred parameters have, when a microphysics constraint is used. To obtain a better measure for the uncertainties when scatter is accounted for, we have artificially increased the error bars of all the data by a factor of $5$ and re-calculated them for the best-fit model of the \textit{k free} class. The results are presented in the bottom row of Table \ref{970508_table}. The choice of the factor is motivated by the value of $\chi^2/\textrm{dof}\approx 1$ that it results in, producing a statistically `good' fit. However, scatter is not the only reason for the high values of $\chi^2/\textrm{dof}$, there are also systematic deviations from the data (for example in the $R$ and $V$ band during the first $20$ days). Therefore, strictly speaking, the method of artificially increasing the error bars should not be applied to the whole data set. Nevertheless, its application results in uncertainties that represent better the parameter range allowed by the data and is not used to draw any conclusions on the quality of the fits.

X-ray data show a preference for the \textit{ISM} class, but hardly influence the fit at all, due to the small number of data points. We have investigated the dependence of our results (especially those for $k$) on the relative importance of X-ray data by increasing the error bars by a factor of 5 in all bands, apart from X rays, and recomputing the uncertainties in the values of the inferred parameters. The best-fit results are essentially identical to those of Table \ref{970508_table} for the \textit{k-free} class. The $1\sigma$ uncertainties, while larger than those presented in the lowest row of Table \ref{970508_table}, exclude the \textit{ISM} scenario.

A discussion of the spectra, dynamics and inferred parameters in the \textit{Wind} scenario (that produces the best fits) is presented in Section \ref{970508_section}.

\begin{table*}
\centering  
\parbox{1.5\columnwidth}{\captionof{table}{Best-fit parameters of each class for GRB\,980703. For column description see Table \ref{970508_table}. \label{980703_table}}}
\small\addtolength{\tabcolsep}{+3pt}
\begin{tabular}{l c c c c c c c c} 
\toprule
 $\textrm{Class}$ & $\textrm{Constraint}$& $E_{52}$ & $n_0$ & $p$ & $\epsilon_{\textrm{B}}$ & $\epsilon_{\textrm{e}}$ & $k$ & $\chi^2/\textrm{dof}$ \\ [0.5ex] 
\hline
\hline
$\textit{ISM}$ & $-$ & $17.82$ & $760.3$ & $2.538$ & $10^{-7}$ & $1.0$ & $0$ & $15.56$ \\ [1.5ex]
\hline
$\textit{Wind}$ & $-$ & $1.771$ & $14.220$ & $3.865$ & $0.0377$ & $0.133$ & $2$ & $9.090$ \\ [1.5ex]
\hline
$\textit{k free}$ & $\textrm{Equipartition}$ & $2.546$ & $4.265$ & $3.933$ & $0.115$ & $0.115$ & $1.154$ & $\textbf{8.691}$  \\ [0.5ex]
\bottomrule\\ [0.1ex]
\end{tabular}
\smallskip
\end{table*}

\subsection{GRB\,980703}
Another well-sampled afterglow that has been extensively modelled in the literature is that of GRB\,980703. We have performed fits to the full data set, from radio to X rays, and we have found that no set of parameters can fit the data. The \textit{Wind} model does better than the \textit{ISM}, but the best fit is obtained for $k\approx1.15$.

In Fig. \ref{980703_lcs} we present light curves from best-fit models of all classes. In the radio, the \textit{ISM} model underperforms compared to the other classes. In the optical and near-infrared none of the models seems to be able to reproduce the data adequately, especially in the low-energy bands. X-ray data, on the other hand, can only be described within the \textit{ISM} class. From this general picture we can conclude that the physical scenario of synchrotron radiation from a spherical blast wave is not realistic for this source.

For every class of models, we have selected those with the microphysics settings that produced the best fits and present them in Table \ref{980703_table}. For the \textit{ISM} and \textit{Wind} class, the model that performs better is the one with no constraint on the microphysics, whereas when $k$ is free, equipartition produces the best $\chi^2/\textrm{dof}$. Fitting the afterglow of this burst we have allowed for $p$ to range between $2.0$ and $4.0$ because requiring $p<3.5$ results in values on the edge of the parameter space. Both the best-fit model of the \textit{k free} class and the one from the \textit{Wind} class have very high values for $p$ ($>3.8$). Their $\chi^2/\textrm{dof}$ values are notably better than those of the \textit{ISM} class. The values of $t_{\textrm{NR}}$ are $100$, $1310$ and $880$ days for the \textit{ISM}, \textit{Wind} and \textit{k free} class, respectively. Due to the overall-bad fits to the light curves and the extreme best-fit values of $p$, we consider the values we obtain unreliable. For that reason we have not calculated any errors on the derived parameters for this burst.

It is worth noting the consensus over the outflow geometry of GRB\,980703. Several studies infer small opening angles and jet breaks in the timescale of days-weeks \citep{Panaitescu2001:ApJ554,Yost2003:apj597,Frail2003:apj590}. In the spherical model, the very fast decays observed in the $H$, $J$ and $R$ bands, can only be explained by very large values of $p$, that result in steep light-curve profiles. However, a more natural explanation of the observed slopes would be that the edge of the jet has become visible \citep{Rhoads1999:Apj525, Panaitescu2005:MNRAS362}. We therefore regard the results presented in this paper implicit confirmation of the jet geometry in the outflow of GRB\,980703.

\begin{table*}
\centering  
\parbox{1.5\columnwidth}{\captionof{table}{Best-fit parameters of each class for GRB\,070125. For column description see Table \ref{970508_table}. \label{070125_table}}}
\small\addtolength{\tabcolsep}{+2pt}
\begin{tabular}{l c c c c c c c c} 
\toprule
 $\textrm{Class}$ & $\textrm{Constraint}$& $E_{52}$ & $n_0$ & $p$ & $\epsilon_{\textrm{B}}$ & $\epsilon_{\textrm{e}}$ & $k$ & $\chi^2/\textrm{dof}$ \\ [0.5ex] 
\hline
\hline
$\textit{ISM}$ & $-$ & $6.968$ & $1.053\!\cdot\!10^3$ & $3.203$ & $1.12\!\cdot\!10^{-5}$ & $1.0$ & $0$ & $14.30$ \\ [1.5ex]
\hline
$\textit{Wind}$ & $-$ & $11.85$ & $1.478\!\cdot\!10^3$ & $2.717$ & $1.59\!\cdot\!10^{-6}$ & $1.0$ & $2$ & $10.80$ \\ [1.5ex]
\hline
$\textit{k free}$ & $-$ & $15.32$ & $3.062\!\cdot\!10^3$ & $2.831$ & $5.19\!\cdot\!10^{-7}$ & $1.0$ & $1.670$ & $\textbf{10.49}$  \\ [0.5ex]
\bottomrule\\ [0.1ex]
\end{tabular}
\smallskip
\end{table*}

\begin{figure*}
\centering
\includegraphics[trim=0cm 0.0cm 0cm 0cm, clip=true, width=1.8\columnwidth]{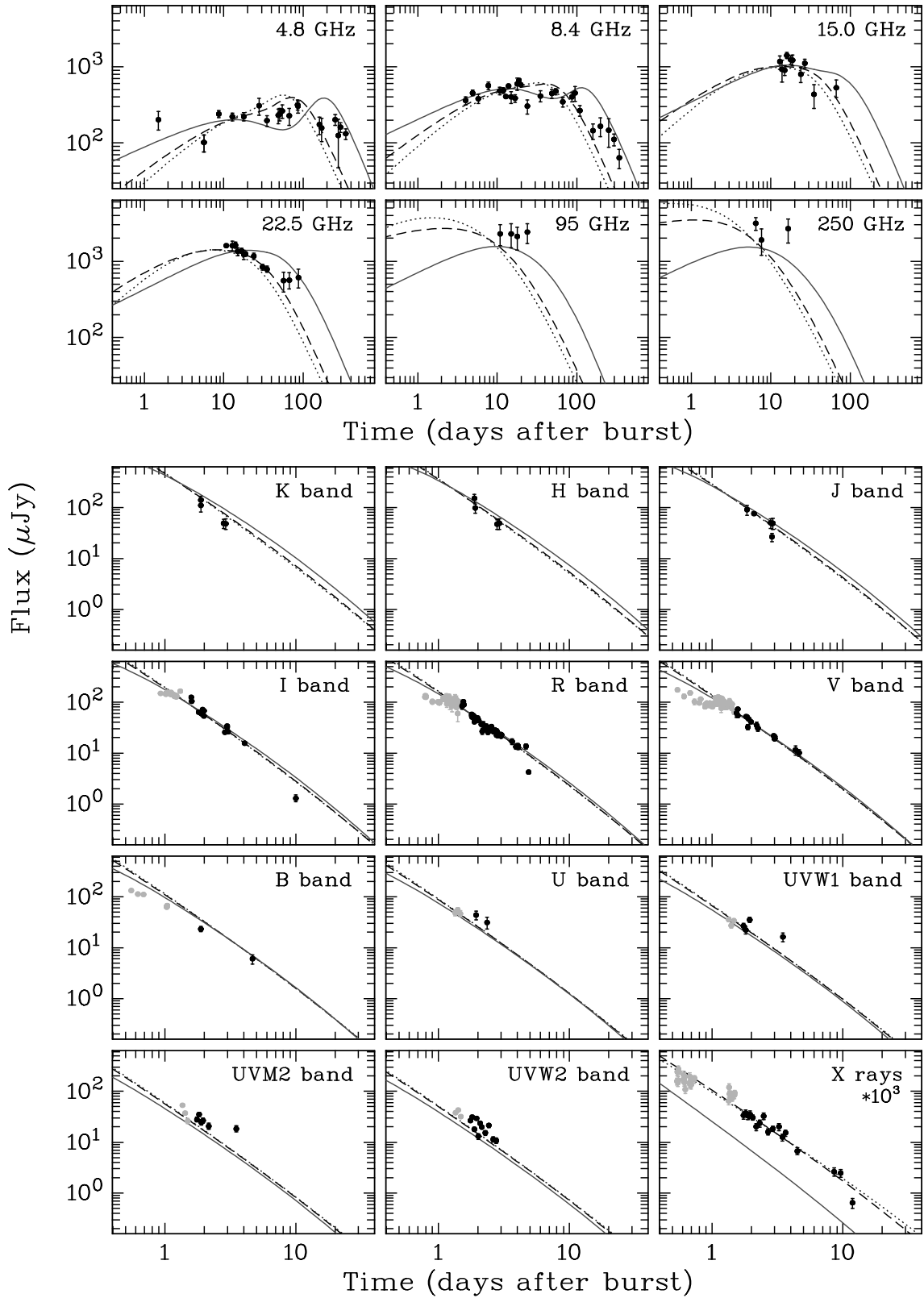}
\parbox{1.8\columnwidth}{\caption{Afterglow of GRB\,070125. Best-fit light curves for \textit{ISM} (solid grey line), \textit{Wind} (dotted black line) and \textit{k free} (dashed black line) classes. Radio and millimetre bands are on top. The lower panel shows near-infrared, optical, ultraviolet and X-ray bands. Data points before $1.5\, \textrm{days}$ have been excluded from the fits but appear in grey colour in the figure. In all bands, data points have $1\sigma$ errors. \label{070125_lcs}}}
\end{figure*}

\pagebreak

\subsection{GRB\,070125}

The afterglow of the exceptionally luminous GRB\,070125 was observed in several bands, lasting more than ten days in X rays and about a year in the radio. We find that \textit{Wind}-like models provide the best description of the data, but with noticeable outliers and with inferred parameters that are fairly extreme ($E>10^{53}\, \textrm{erg}$, $\epsilon_{\textrm{e}}=1$, $\epsilon_{\textrm{B}}<10^{-5}$). We consider the results indicative, but by no means conclusive, as additional physics (e.g. jets) may be needed to explain the deviations and special conditions are required to account for the physical parameters we obtain.

In Fig. \ref{070125_lcs} we present light curves of the best-fit models from each class. Results for the \textit{Wind} and \textit{k free} classes are similar to each other and differ significantly from the \textit{ISM} class in radio, millimetre and X-ray bands, where the former perform better. However, late-time behaviour of the data at $4.86,\,8,46$ and $22.5\,\textrm{GHz}$, as well as millimetre observations are hard to explain within any model. In the near-infrared and optical bands all classes produce good fits. In the ultraviolet, there is a slight underestimation of the flux levels. In X rays, only \textit{Wind} and \textit{k free} models are able to describe the data.

In agreement with \cite{DeCia2011:MNRAS418} we find that $\nu_{\textrm{c}}$ lies between optical and X-ray energies throughout the observations. \cite{Chandra2008:ApJ683}, on the other hand, find that they can best explain the data when $\nu_{\textrm{c}}$ lies below the optical. Calculation of $t_{\textrm{NR}}$ yields $80$ days in the \textit{ISM} case and $\sim140$ days in the other classes. This ensures that the relativistic formula for $\nu_{\textrm{c}}$ is valid during near-infrared, optical, ultraviolet and X-ray observations, that last up to 10 days after the gamma-ray trigger. In all classes, $\nu_{\textrm{m}}$ starts off bellow the optical and overtakes $\nu_{\textrm{a}}$ at $30-80$ days. The different temporal evolution of $\nu_{\textrm{a}}$ makes for the deviations in late radio light curves between the \textit{ISM} class and the others.

In Table \ref{070125_table} we present the values of the inferred parameters for the best-fit models of each class. Deviations between different classes are moderate. Best $\chi^2/\textrm{dof}$ values are found when no assumption for the microphysics is made. This is because to explain the data, all models require a high value for $\epsilon_{\textrm{e}}$ and a very low one for $\epsilon_{\textrm{B}}$. Values of the parameters when $k$ is free (model with the best $\chi^2/\textrm{dof}$) are closer to those from the \textit{Wind} scenario, without, however, matching them. The inferred energies are high in all cases, as are the values for $p$. Given the imperfect fits and the extreme parameters we infer, we have not calculated errors for their values.


\section{The curious case of GRB\,970508}\label{970508_section}

GRB\,970508 is a unique burst in many ways. Its afterglow was only the second ever observed and despite the multi-frequency monitoring, in some bands over the period of several months, the inferred physical parameters vary widely between difeerent authors \citep{Wijers1999:apj523, Chevalier2000:ApJ536, Frail2000:apj537, Panaitescu2002:ApJ571, Yost2003:apj597, Berger2004:ApJ612}. From our sample, the fits to GRB\,970508 are deemed the most reliable and the most successful, despite the higher values for $\chi^2/\textrm{dof}$. There are two basic reasons for this. The first is the overall behaviour of model light curves that successfully reproduce the trends of the data at all well-probed wavelengths. The second reason is the stability and convergence that the fits to GRB\,970508 show, especially in the \textit{Wind} scenario, but also when $k$ is a free parameter and a constraint on the microphysics is imposed. If no constraint is placed on the microphysics and $k$ is a free parameter, we cannot discern between the \textit{ISM} and the \textit{Wind} scenario. However, once either equipartition or the Medvedev formula are used, the results clearly favour a wind-type CBM. In this Section we present an analysis of the physics implied by the best-fit parameters we obtain for GRB\,970508 and compare those to values inferred by other authors.

\subsection{Microphysics}
 At first glance, the best-fit values presented in Table \ref{970508_table} reveal an issue concerning the microphysics of the blast-wave, namely, the sum of $\epsilon_{\textrm{B}}$ and $\epsilon_{\textrm{e}}$ is greater than 1. In fact, in order for the outflow to be adiabatic, as assumed by the model, at least one of these parameters has to be much smaller than 1. A low value of $\epsilon_{\textrm{e}}$ ensures that most of the energy remains in the blast wave, even if the electrons radiate efficiently, while a low value of $\epsilon_{\textrm{B}}$ moderates the energy losses of the electron population. The degeneracy of the theoretical model \citep{Eichler2005:apj627} which has prompted us to freeze $\xi=1$ during the fit process, can be used to solve this issue. The net effect of this degeneracy is that a set of parameters $E_{52}'=f^{-1}\,E_{52},\,n_0'=f^{-1}\,n_0,\,\epsilon_{\textrm{e}}'=f\,\epsilon_{\textrm{e}},\,\epsilon_{\textrm{B}}'=f\,\epsilon_{\textrm{B}},\,\xi'=f\,\xi$ produce the same spectrum as the unprimed ones, regardless of the value of (the positive number) $f$. Therefore, the inconsistency implied by the high values of $\epsilon_{\textrm{B}}$ and $\epsilon_{\textrm{e}}$ may be seen as evidence that $\xi<1$, which means that \textit{not all electrons are accelerated at the shock}. Consequently, the values for $E_{52}$ and $n_0$ presented in Table \ref{970508_table} should be viewed as lower limits, whereas those for $\epsilon_{\textrm{B}}$ and $\epsilon_{\textrm{e}}$ as upper limits.

Another notable feature of the results for the microphysics in the \textit{Wind} scenario is that we can not conclusively distinguish between the three possibilities (no constraint, equipartition, Medvedev relation). Equipartition settings seem to be marginally favoured by the better $\chi^2/\textrm{dof}$ values these models have, but the Medvedev relation cannot be ruled out. The ambiguity of our results is mainly caused by the relatively high values that both $\epsilon_{\textrm{B}}$ and $\epsilon_{\textrm{e}}$ have. We have run fits where $\xi$ was frozen at $0.1$ and $0.01$ and monitored the behaviour of the two former quantities. They were found to be approximately equal to each other and always (as did $E_{52}$ and $n_0$) followed the scalings implied by the degeneracy relations. This confirms energy equipartition between power-law electrons and magnetic field, which is also suggested by $\chi^2/\textrm{dof}$ values.

\subsection{Spectra}
In the \textit{Wind} scenario the synchrotron spectrum starts off at $1.5$ days exhibiting fast cooling \citep{Sari1998:apjl497} with the critical frequencies having the following values: $\nu_{\textrm{a}}=1.2\cdot 10^{10}\, \textrm{Hz}$, $\nu_{\textrm{c}}=1.1\cdot 10^{12}\, \textrm{Hz}$, $\nu_{\textrm{m}}=1.4\cdot 10^{13}\, \textrm{Hz}$. At $5$ days, $\nu_{\textrm{m}}$ overtakes $\nu_{\textrm{c}}$, causing the wiggle in the radio light curves of the model (see Fig. \ref{970508_lcs}). The flux at the highest-frequency power law of the spectrum (where near-infrared, optical, ultraviolet and X-ray data lie) is independent of the ordering of critical frequencies and therefore no feature is observed in those bands during the spectral transition. After $5$ days the spectrum settles into the slow-cooling regime. During the fast-cooling phase (i.e. before $5$ days), almost all available data lie above $\nu_{\textrm{m}}$ and $\nu_{\textrm{c}}$; there are hardly any significant radio observations during that time. Therefore, our approximations for $\nu_{\textrm{a}}$ when $\nu_{\textrm{a}}<\nu_{\textrm{c}}<\nu_{\textrm{m}}$ have a negligible effect on the fits. Moreover, given that the values of $\nu_{\textrm{m}}$ and $\nu_{\textrm{c}}$ are largely independent of their ordering in the spectrum \citep{Granot2002:Apj568, vanEerten2009:mnras394}, the validity of our approach towards optical data is ensured. From $5$ days onwards, no approximation is made for the value of any of the critical frequencies and the model assumes its most accurate form.

It is worth noting that the best-fit spectra naturally explain the spectral evolution (at $\sim100$ days) depicted in Fig. 5 of \cite{Frail2000:apj537}, due to the passage of $\nu_{\textrm{m}}$. In the \textit{ISM} case $\nu_{\textrm{m}}$ crosses the radio earlier, at around $45$ days, something excluded by the data. \cite{Frail2000:apj537} also find $\nu_{\textrm{a}}=3\,\textrm{GHz}$ at seven days, whereas in our best fit $\nu_{\textrm{a}}=5.5\,\textrm{GHz}$, at the same observer time. We consider the difference negligible, especially considering the strong variation the light curves show around those observer times, due to scintillation. This can be verified by inspection of Fig. 4 of \cite{Frail2000:apj537}, where the spectral index between $4.86$ and $8.46\,\textrm{GHz}$ varies between $0.4$ and $1.6$ within the first two weeks. We have also checked the claim of \cite{Galama1998:ApJ500L} who suggest that $\nu_{\textrm{c}}$ is observed to pass through the near-infrared bands at $\sim 10$ days. When all the available data from several different bands ($K,\,I,\,R,\,V$) are taken into account, we find that the spectral index starts off (at $\sim2$ days) having values consistent with late time observations, thus showing no evidence of spectral evolution.

In the best-fit model of the \textit{Wind} scenario, $\nu_{\textrm{c}}$ lies below the optical bands throughout the duration of optical observations. Therefore, its exact value is important at all observer times. As mentioned in Section \ref{970508_sub}, calculation of $t_{\textrm{NR}}$ yields $\sim 180$ days. This implies that the values of $\nu_{\textrm{c}}$ during late near-infrared and optical observations (extending up to $\sim 200$ days in the $R$ band) should be mildly affected by the transition towards the Newtonian dynamical phase. Since $\nu_{\textrm{c}}$ is not included in the treatment of \cite{Leventis2012:MNRAS427} we do not have a description of the transition for this critical frequency, at least not at the level of accuracy that we do for the others. Assuming that the trans-relativistic behaviour of $\nu_{\textrm{c}}$ is similar to those of the other spectral parameters (smoothly broken power-law) and that the break is centered around $t_{\textrm{NR}}$, we have explored various sharpnesses for that transition and found that the fit results remain consistent. The only parameter that changes noticeably is $p$ which grows from $2.28$ to about $2.34$ when the transrelativistic evolution of $\nu_{\textrm{c}}$ is taken into account. Having established that this evolution does not affect the inferred parameters, the results we present in Table \ref{970508_table} are obtained using the relativistic formula for $\nu_{\textrm{c}}$ only.

\subsection{Transrelativistic phase}
First noticed in \cite{vanEerten2010:mnras403} and subsequently quantified in \cite{Leventis2012:MNRAS427}, the duration of the trans-relativistic phase of a spherical outflow in the observer frame can be long (this also holds in the case of a jetted outflow; \citealt{Zhang2009:Apj698}). The near-infrared and optical light curves in Fig. \ref{970508_lcs} show strong deviations from the ultrarelativistic behaviour already at a few tens of days, in observer time. Their progressive steepening is caused entirely by the dynamics slowly adjusting to the Sedov-Taylor solution, as there is no critical frequency crossing these bands. The effect is similar in the radio, but less pronounced due to the simultaneous spectral evolution.

Deviations of the observed radio light curves from the relativistic scalings at timescales of several weeks prompted \cite{Waxman1998:ApJ497} to propose a jetted outflow for GRB\,970508. In this paper we demonstrate how accurate modelling of the transrelativistic phase can account for the deviations from the ultrarelativistic scalings at observer times $\ll t_{\textrm{NR}}$. This implies that a similar trend may hold for at least some other GRB afterglows, the temporal evolution of which has been interpreted as evidence for a jet break. 

Differentiating between jet breaks and the transition to the non-relativistic phase is important, as it directly affects the inferred geometry and energetics of GRB outflows. There are two main quantities that can serve as diagnostics for this differentiation. The first is the change of the temporal index of the flux. In the case of a jet break, a decrease in the value of the temporal index is expected, mainly due to the missing-flux effect \citep{Panaitescu1998:ApJ503} that arises when the edges of the jet become visible. On the other hand, the change in the temporal slope, as the outflow approaches the transition to non-relativistic velocities, may be positive or negative, is a function of $k$ and (depending on the spectral regime) $p$, and is known from theory (e.g. \citealt{vanEerten2010:mnras403}). A second diagnostic is the duration (smoothness) of the change in the temporal index. In the case of a jet break the transition lasts from factors of few (ISM environment) up to a decade (wind environment) in observer time \citep{DeColle2012b, vanEerten2012:arXiv1209}, whereas the typical duration of the transrelativistic regime in the case of spherical outflows is a few decades \citep{Leventis2012:MNRAS427}. The picture is slightly more complicated in the case of a jet break observed off axis. In that case the jet-break transition is effectively stretched and postponed (in reality it splits in two). Good coverage is then critical to discern between the different interpretations.

\subsection{Comparison to previous work}
Several broadband fits to the afterglow of GRB\,970508 have been performed and presented in the literature \citep{Wijers1999:apj523, Chevalier2000:ApJ536, Yost2003:apj597, Panaitescu2002:ApJ571}. Others have fit only late-time radio data \citep{Frail2000:apj537, Berger2004:ApJ612}, while \cite{Starling2008:apj672} have fit only the slopes of light curves and spectra to infer values for $p$ and $k$. Most of these studies assume or find that an \textit{ISM} scenario fits the data better, apart from \cite{Chevalier2000:ApJ536} and \cite{Panaitescu2002:ApJ571} who favour the \textit{Wind} case. In this study we have presented a detailed investigation of both density structures that clearly favours a stellar-wind CBM. In addition we demonstrate how models with no assumption on the slope of the CBM converge to the \textit{Wind} scenario. Interestingly, \cite{Starling2008:apj672} find that, in their sample, four out of five afterglows with well-constrained values for $k$ suggest the same. In that study the density structure of GRB\,970508 is poorly constrained.

There seems to be more agreement on the geometry of the outflow of GRB\,970508. Most studies (also \citealt{Rhoads1999:Apj525}) do not need to invoke a jet, while those that do infer a jet geometry, usually find large half-opening angles: $18^{\circ}$ \citep{Panaitescu2002:ApJ571}, $30^{\circ}$ \citep{Frail2000:apj537}, $50^{\circ}$ \citep{Yost2003:apj597}. We find that the spherical model provides a good description of the data, capturing the trends of the light curves at different wavelengths for more than two orders of magnitude in observer time. We argue that GRB\,970508 may have indeed originated from an almost spherical outflow. The energy of the prompt emission is estimated around $5\cdot 10^{51}\, \textrm{erg}$, if isotropic \citep{Bloom2001:AJ121}. Although on the high side, this value is not unreasonable (e.g. \citealt{Metzger2011MNRAS413}).

In terms of the whole set of fitted parameters, our results are similar to those of \cite{Chevalier2000:ApJ536} and \cite{Panaitescu2002:ApJ571}. Given the uncertainties in the last row of Table \ref{970508_table}, their best-fit values are within, or just outside the allowed range of our results. We find moderately higher values for $\epsilon_{\textrm{e}}$ and $\epsilon_{\textrm{B}}$ than both studies, but these values are effectively upper limits. Lowering $\xi$ to $0.3$ results in $E_{52}\simeq 0.4$, $A_{\ast}\simeq 0.73\, \textrm{g cm}^{-1}$, $\epsilon_{\textrm{e}}\simeq\epsilon_{\textrm{B}}\simeq0.19$, while the value for $p$ remains the same, $2.28$. None of these parameters are more than a factor of three off compared to both aforementioned studies (note, however, the inference of a jet from \citealt{Panaitescu2002:ApJ571}). It is worth mentioning that the blast-wave energy inferred through modelling of the afterglow radiation is very similar to the radiative output of the prompt emission. This result holds regardless of the outflow geometry and implies a very high efficiency of the gamma-ray radiation from the main burst. However, given the fast cooling at early times, adiabatic evolution of the blast wave demands $\epsilon_{\textrm{e}} \ll 1$. For $\xi<0.17$, both $\epsilon_{\textrm{e}}$ and $\epsilon_{\textrm{B}}$ are smaller than $10\%$. The corresponding blast wave energy becomes $>8 \cdot 10^{51}\, \textrm{erg}$, which reduces the efficiency of the prompt emission below $40\%$.


\section{Discussion}\label{discussion}

In this Section we discuss the implications of our results for the properties of GRB outflows and afterglow fitting.

\subsection{Collimation of GRB outflows}
We have demonstrated how a spherical outflow can account for the observations of GRB\,970508 and how it fails in the case of GRB\,980703. The former afterglow has often been successfully modelled both with a spherical and a collimated outflow. For GRB\,980703 a jet is invariably inferred and in this research, similarly to \cite{Frail2003:apj590}, we find that a spherical model cannot provide an adequate description to the data under any combination of physical parameters.

The degree of collimation in the case of GRB\,070125 is less clear. On the one hand, there are only a few studies of the afterglow radiation and only one of them performs broadband (radio to X rays) fitting \citep{Chandra2008:ApJ683}. On the other hand, our results provide a satisfactory description of most of the broadband data, apart from late time behaviour in the radio, when an additional component is observed in the light curves. This component, however, cannot be explained in the jetted model of \cite{Chandra2008:ApJ683} either. Moreover, they propose that the X-ray flux is dominated by inverse Compton, in order to explain what seems to be a \textit{chromatic} break in the optical and X-ray light curves ( at about $4$ and $10$ days, respectively). However, the claim for a jet-break in the optical is based only on two data points (one in the $I$ and one in the $R$ band), while in the X rays it is only based on one (see Fig. \ref{070125_lcs}). We find that a spherical model offers a similar level of accuracy, without the need to invoke a jet or other radiation mechanisms beyond synchrotron. However, the parameters we obtain are extreme, both on the microphysics side, but also in the total energy budget they imply ($>10^{53}\,\textrm{erg}$). We therefore consider it likely that the model we have used lacks some physics, which at least in some bands and observer times `drives' the radiated spectrum. That extra physics could be a jetted outflow, but the evidence from previous studies combined with our findings is not conclusive.

In this study we cannot quantify the opening angle of jets, in cases that one is inferred. We can, however, qualify afterglows as spherical by successfully fitting their broadband data set. This has been the case for GRB\,970508 and we consider this a clear demonstration of the diversity in the geometry of GRB outflows. This is in accordance with searches for jet breaks in large samples of afterglow observations that fail to clearly identify a jet break in more than half of the sources \citep{Kocevski2008:ApJ680, Racusin2009:ApJ698}. However, in the collapsar model \citep{MacFadyen1999:ApJ524} for long GRBs, it is likely that outflows are still collimated right after breaking out of the stellar envelope (e.g. \citealt{Morsony2007:ApJ665}). If the opening angle of the jet is large, the light curves will exhibit deviations from spherical symmetry during the transrelativistic phase. In such a quasi-spherical scenario, the observational signatures of decollimation might be weak and the expected differences from a perfectly spherical outflow have, to the best of our knowledge, not been explored in the literature. In the case of GRB\,970508, $t_{\textrm{NR}}=180\,\textrm{days}$ in the \textrm{Wind} class, and the decollimation should occur on similar timescales, close to the end of data sampling. Therefore, if the outflow of GRB\,970508 had a very large opening angle, our fits will be dominated by observations of an almost conical flow. The inferred energy would then be the isotropic equivalent of the real energy content of the blast wave, which is lower only by a factor of order unity. The best-fit values of $p$ and $k$ are inferred by the slopes of spectra and light curves which, at least for the best part of the observations, are not influenced by effects caused by a possible quasi-spherical geometry. Therefore, we would not expect our general conclusions concerning the slope of the CBM to be significantly affected by such a scenario.

Quantifying the distribution of jet opening angles is not an easy task, especially considering the inadequate (for broadband modelling) coverage that a large fraction of afterglows have. On the observational side, \cite{Curran2008:MNRAS386} have shown that jet breaks may be misidentified as single power laws, due to data-analysis effects. Moreover, \cite{vanEerten2010:Apj722} have shown that a moderately off-axis viewing angle (but smaller than the jet semi-opening angle) can `mask' the appearance of a jet-break. If jets are present, observing them off axis should happen more often than not. Therefore, this is an important effect that should be taken into account in the model fits. Another issue that needs to be better understood is the early ($10^3-10^5\,\textrm{s}$) afterglow behaviour which in a large fraction of bursts suggests some form of energy injection, continuous or irregular \citep{Nousek2006:ApJ642, Panaitescu2011:MNRAS414}. This may affect the overall dynamics of the outflow but also result in misinterpreting a potentially coincident jet break \citep{Racusin2009:ApJ698}. Thus, connecting the dynamics of the early afterglow with the more regular behaviour observed at larger timescales is essential to uncover evidence for jets that may not be in the form of the canonical achromatic jet break.

\subsection{The immediate environments of gamma-ray bursts}
In this work we have treated the density structure of the CBM as a free parameter ($k$), assuming that a constant power law applies. Out of the three data sets we studied, one (GRB\,970508) showed convergence to a constant stellar wind, represented by $k=2$. The best fit to GRB\,980703 is obtained for $k=1.154$. Lastly, for GRB\,070125 the best-fit value of $k=1.67$, which is closer to that of a constant stellar wind than homogeneous CBM. For all data sets, \textit{Wind} environments produce better fits than the \textit{ISM} class.

Similarly, however, to the discussion on the geometry of the outflows, GRB\,970508 is the only one with reliable results. In both GRB\,980703 and GRB\,070125 large values for both $p$ and $k$ are needed to best describe the data within the spherical model, the applicability of which is at least doubtful in these cases. For GRB\,970508, the value of $A_{\ast}$ implies that the inferred density profile corresponds to a constant mass-loss rate of $2.4\cdot10^{-6} \, \xi^{-1} \, \,M_{\odot}/\textrm{yr}$, for a wind velocity of $1,000\,\textrm{km}\,\textrm{s}^{-1}$. Interpreting the adiabatic condition as $\xi<0.17$, we find $\dot{M} > 1.5\cdot10^{-5} \,M_{\odot}/\textrm{yr}$, which implies a relatively massive Wolf-Rayet star towards the end of its life \citep{Chevalier1999:ApJ520L}.

Several studies have fit individual bursts and found or assumed a homogeneous density structure for the CBM. When the fits are compared against those with stellar-wind CBM the results are often ambiguous (e.g. \citealt{Frail2003:apj590}; \citealt{Chandra2008:ApJ683}), while in some cases the \textit{Wind} scenario seems to be favoured \citep{Chevalier2000:ApJ536,Panaitescu2002:ApJ571}. On the theoretical side, \cite{vanEerten2012:arXiv1209} have shown that the majority of \textit{Swift} post jet break slopes are not reconcilable with a constant density CBM, if late energy injection and viewing angle do not significantly affect the observations. Instead, the observed slopes suggest a wind-type environment for the CBM. \cite{Starling2008:apj672} have studied a sample of 10 \textit{Beppo-SAX} afterglows and found that the majority of the data sets that were sufficient to constrain the value of $k$ implied a stellar-wind CBM. However, half of them have error bars that allow for a wide range for $k$. \cite{Curran2009:mnras395} have performed a similar study using \textit{Swift} bursts and find a division in the sample between constant and wind-like profiles. It seems, therefore, likely that the density structure of the CBM in GRBs is diverse, similar to the geometric characteristics of their outflows. However, this does not necessarily translate to diversity of the progenitors as well, because in the collapsar model \citep{Woosley1993:Apj405, MacFadyen1999:ApJ524} the CBM of a large fraction of long GRBs is modified by multiple stellar winds from the neighbouring stars \citep{Mimica2011:MNRAS418}.

\subsection{Model constraints}
For all the afterglows we studied, we have found that a multi-frequency data set is more suitable for fitting all the parameters at once. This has led to the expansion of the model with the inclusion of the cooling frequency of the synchrotron spectrum, $\nu_{\textrm{c}}$. However, even when radio to X-ray data are fitted and all details of the spectrum are taken into account, setting $k$ a free parameter results in large uncertainties, if no assumption for the microphysics is made. This is manifested in the large errors for the best-fit values of physical parameters in the case of GRB\,970508 (see row 7 of Table \ref{970508_table}). 

When $k$ is free and no microphysics assumption is made, the number of fitted parameters is six, equal to the maximum number of constraints we can have from the light curves -- four from the positions of the critical frequencies and the value of $F_{\textrm{m}}$, plus two more from the slopes of spectra and light curves. However, our results imply that not all of these constraints are efficiently used during the fitting process. This means that the effects of two or more of the constraints cannot be separated, leading to a case-specific degeneracy. In the case of GRB\,970508, for the best-fit model, both $\nu_{\textrm{m}}$ and $\nu_{\textrm{c}}$ lie between radio and near-infrared bands for the best part of the observations ($\nu_{\textrm{m}}$ stays above the radio bands for about $100$ days). Therefore, their positions are not independently constrained by the data, leading to a wide range of possible values when all six parameters are simultaneously fitted.

An interesting feature of the prescriptions we have used is the inclusion of $\xi$ as a parameter. Due to the degeneracy of the model, the presence of $\xi$ is not necessary per se. One can imagine a situation where a range in the allowed values for $\xi$ is reflected in the adjustment of the ranges of the other parameters. For example, by assuming that $\xi=1$ and allowing $\epsilon_{\textrm{B}}$ and $\epsilon_{\textrm{e}}$ to obtain values $>1$ during fitting, one accounts for the possibility of $\xi$ being smaller than those two parameters, while all of them are smaller than unity. However, the inclusion of $\xi$ in the model demonstrates these situations more clearly. In the results we obtain for GRB\,970508 it was not initially possible to discern between the Medvedev constraint and the equipartition constraint for the microphysics due to the high values of both $\epsilon_{\textrm{B}}$ and $\epsilon_{\textrm{e}}$, that, within the uncertainties, extend to the upper limit of the allowed range. By freezing $\xi$ at values much lower than 1, we have excluded the presence of better fits in which $\epsilon_{\textrm{B}}>\xi$ and/or $\epsilon_{\textrm{e}}>\xi$, and confirmed that energy equipartition between power-law electrons and magnetic field describes better the afterglow observations of GRB\,970508.


\section{Conclusions}\label{conclusions}

We have performed broadband fits of three afterglow data sets using accurate analytic flux prescriptions applicable to spherical outflows. We have shown that GRB\,970508 is successfully fit by a spherical model. The fits fail in the case of GRB\,980703 and GRB\,070125 at varying degrees, implying that these sources may be indeed related to jetted outflows. This is supported by extensive modelling of the former and the extremely high isotropic energy inferred for the latter.

For GRB\,970508 we find that the best-fit value for $k$ is practically 2, strongly suggesting a stellar-wind environment. Fits to GRB\,970508 also show strong evidence for a population of electrons that is not accelerated at the forward shock. The implied values for the microphysics parameters, $\epsilon_{\textrm{e}}$ and $\epsilon_{\textrm{B}}$, suggest that they are close to equipartition.

Modelling of GRB\,970508 illustrates how an accurate spherical model accounts for the progressive deviations of light curves from the ultrarelativistic scalings at $t_{\textrm{obs}} \ll t_{\textrm{NR}}$. This feature had been previously interpreted as a jet break in the context of simpler models, but emerges naturally from precise calculations of dynamics and spectra in the spherical scenario. Therefore, we consider it possible that similar features in the data sets of other afterglows have been misinterpreted as jet breaks, in the absence of detailed calculations for the spherical case.

\section{Acknowledgements}

We would like to thank A. De Cia for providing the data for GRB\,070125. This research was supported by NOVA and in part by NASA through grant NNX10AF62G issued through the Astrophysics Theory Program and by the NSF through grant AST-1009863. RAMJ and AJvdH acknowledge support from the ERC via Advanced Investigator Grant no. 247295. We thank SARA Computing and Networking Services (www.sara.nl) for their support in using the Lisa Compute Cluster.

\bibliographystyle{mn2e}
\bibliography{Fits_spherical}

\label{lastpage}

\end{document}